\PassOptionsToPackage{table}{xcolor}
\documentclass[sigconf,natbib=false,nonacm]{acmart}
\AtBeginDocument{%
  }

\usepackage{amsmath,amsfonts}
\usepackage{algorithmic}
\usepackage{graphicx}
\usepackage{textcomp}
\usepackage[style=acmnumeric]{biblatex}
\usepackage{cleveref}
\usepackage{xcolor}
\usepackage{microtype}
\usepackage[utf8]{inputenc}

\setcopyright{cc}
\setcctype{by}
\acmDOI{10.1145/3832783.3834527}
\acmYear{2026}
\copyrightyear{2026}
\acmISBN{979-8-4007-2882-2/2026/10}
\acmConference[ASE '26]{Proceedings of the 41st IEEE/ACM International Conference on Automated Software Engineering}{October 12--16, 2026}{Munich, Germany}
\acmBooktitle{Proceedings of the 41st IEEE/ACM International Conference on Automated Software Engineering (ASE '26), October 12--16, 2026, Munich, Germany}
\acmSubmissionID{ase26ind-p344-p}
\received{2026-05-01}
\received[accepted]{2026-07-01}

\begin{document}

\title{From Backlog Items to Security Guidance: Towards Continuous Security Compliance}

\author{Ignacio García Núñez}
\orcid{0000-0002-2174-2703}
\affiliation{%
  \institution{Technical University of Munich}
  \city{Munich}
  \country{Germany}
}
\email{garcia.nunez@tum.de}

\author{Florian Angermeir}
\orcid{0000-0001-7903-8236}
\affiliation{%
  \institution{fortiss}
  \city{Munich}
  \country{Germany}
}
\affiliation{%
  \institution{Blekinge Institute of Technology}
  \city{Karlskrona}
  \country{Sweden}
}
\email{angermeir@fortiss.org}

\author{Fabiola Moyón Constante}
\orcid{0000-0003-0535-1371}
\affiliation{%
  \institution{Siemens Technology}
  \city{Munich}
  \country{Germany}
}
\affiliation{%
  \institution{Technical University of Munich}
  \city{Munich}
  \country{Germany}
}
\email{fabiola.moyon@siemens.com}

\begin{abstract}
Continuous software engineering in regulated domains requires engineering teams to address security throughout the development lifecycle.
Yet making security requirements explicit in backlog items is still problematic.
Engineers must instead infer security relevance of backlog items from brief, free-form descriptions and often lack timely guidance on applicable requirements. We present an NLP-based backlog enrichment system that detects security-relevant backlog items and links them to relevant security requirements. The approach combines a security-relevance classifier with a retrieval-augmented generation (RAG) pipeline over security requirements documents. The approach was developed and evaluated in the context of a large enterprise in highly regulated domains. We present three contributions. First, we release a dataset of 288 backlog items labeled for security relevance by nine security practitioners, with substantial agreement (Fleiss' $\kappa=0.787$). Second, a recall-oriented classifier achieving $F2=0.774$ in-distribution and mean zero-shot G-measure $\approx 0.65$ across five established benchmarks, matching or outperforming most published classical-ML and open-source GPT baselines. Third, we preliminarily evaluated a four-stage security requirements document-grounded RAG pipeline with two practitioners on industrial backlogs using company-internal security policies and CIS Benchmarks. Of the retrieved 24 clauses, 12 were rated at least 4/5 for relevance.
Our findings provide first indicators that NLP-based product backlog enrichment can support engineers in identifying security requirements early in the development process.
With this work we aim to facilitate continuous security compliance through proactive introduction of security requirements in continuous software engineering.
\end{abstract}




\maketitle

\section{Introduction}

Security compliance is no longer confined to a small set of safety- or security-critical domains~\cite{moyon2020security-compliance-agile}.
Regulations such as the EU Cyber Resilience Act~\cite{eu2024cra} and a growing body of sector-specific standards (e.g., IEC 62443 for industrial automation and control system development~\cite{iec2018secure}) are making security a baseline expectation for nearly every organization that builds or operates software.
In organizations leveraging agile or DevOps, this circumstance requires engineering teams to continuously address security throughout the development process~\cite{fitzgerald2017continuous, turpe2017scrum}.
In these settings the product backlog is the central planning artifact~\cite{iso2018sse, ferreira2022writing-agile-requirements}, yet backlog items rarely carry explicit security or compliance annotations~\cite{turpe2017scrum}.
Security considerations are therefore mostly embedded implicitly in short, free-form engineering tasks. This leaves product teams without clear guidance on specific security matters and creates a practical mismatch between continuous software engineering and compliance efforts.
Security planning and reviewing is still often performed manually and outside the main backlog flow, which does not scale to fast development cycles~\cite{moyon2020security-compliance-agile}.
Moyón et al.~\cite{moyon2024challenges} describe this tension in secure continuous software engineering through several open practical challenges, such as making security architecture visible in backlog artifacts and documentation, involving security activities with minimal lead-time burden, and enabling security knowledge and ownership within engineering teams.

This manuscript reports on building, deploying, and evaluating a natural-language backlog enrichment system that addresses this gap directly in the artifact developers already use.
The system surfaces security-relevant context at the product planning stage by attaching security requirements relevant to the organization to flagged backlog items.
The work was developed and evaluated at a large enterprise in highly regulated domains. To enable reproduction, we additionally evaluate the approach on public security guidelines.

The paper makes three contributions:
\begin{itemize}
    \item[\textbf{C1}] An expert-annotated dataset of $288$ backlog items carrying binary security-relevance labels, drawn from two public Atlassian projects and multi-rated by $9$ security practitioners.
    \item[\textbf{C2}] A recall-oriented binary classifier that flags security-relevant backlog items as a first-stage filter for the backlog enrichment system.
    \item[\textbf{C3}] A retrieval pipeline that links backlog items to applicable security requirements from requirement documents.
\end{itemize}

We evaluate the three contributions from complementary perspectives: inter-rater reliability and dataset characterization for C1, in-distribution performance of C2 on C1 and cross-distribution transfer to the Wu et al.~\cite{wu2022data-quality} datasets, and a preliminary evaluation via two practitioner interviews on industrial backlogs for C3, which covers both proprietary and public security policy corpora.

Across the three contributions, a single overarching finding emerges. Security-relevance of backlog items can be evaluated based on the ordinary item text, and concrete security requirements can be attached to those items in a form that engineers and security experts can act on without leaving their existing workflows.

\section{Background}

Three threads of related work are relevant to the contributions of this paper: secure continuous software engineering and the role of the product backlog, detection of security-relevant development artifacts, and the operationalization of normative requirements, such as regulations or standards, into developer guidance.

\subsection{Secure Continuous Software Engineering}

Continuous software engineering shortens delivery cycles and emphasizes incremental, just-in-time work~\cite{fitzgerald2017continuous}.
In contrast, security compliance is traditionally organized as a parallel, formalized track running alongside development, as it is composed of review checkpoints, traceable evidence, and audit-oriented documentation~\cite{moyon2018towardcsc, angermeir2024automatedcsc}.
This parallelism is a practical bottleneck: through case studies in multiple industrial contexts Moyón et al.~\cite{moyon2024challenges} confirmed open industrial challenges blocking adoption at scale, whose consequences share a pattern at the artifact layer.
A lack of item-level security visibility re-introduces compliance checks at release time as a blocking task, security 
activities with high lead time are deferred under pressure and re-emerge as 
security debt, and security knowledge concentrated on experts creates a review 
bottleneck for the rest of the organization.
The contribution of this paper handles these consequences directly: making security relevance visible at the item level (C1, C2) and attaching applicable requirements to backlog items at the point of planning (C3).

\subsection{Product Backlog}

Continuous software engineering teams use product backlogs and issue trackers as their central planning artifact~\cite{iso2018sse}.
Backlog items are typically short, informal, and optimized for coordination rather than for documentation of non-functional concerns, such as security compliance~\cite{ferreira2022writing-agile-requirements}.
As a consequence, when an item carries security implications, those implications are usually embedded implicitly in the free-text description rather than defined as labels, fields, or linked requirements~\cite{raharjana2021user-stories-nlp}, which is the artifact-layer gap that this paper targets.

In this manuscript, we use the definition of Backlog and Backlog Item derived from the ISO/IEC/IEEE 26515:2018~\cite{iso2018sse}: \textit{``A collection of agile features (3.7) or stories of both functional and nonfunctional requirements that are typically sorted in an order based on value priority''}.
Related to that definition is the definition of feature, also provided in that document: \textit{``functional or nonfunctional distinguishing characteristic of a system''}.

Since our focus is on delivering security-relevant information to engineers during planning and development, this manuscript does not differentiate between different backlog types such as product backlogs or sprint backlogs.

\subsection{Security Requirements}
For this work we use the broad term security requirements to refer to requirements related to security formalized in various documents, such as company-internal security policies, security standards (e.g., IEC 62443-4-1), contracts, or public best practice guidelines (e.g., CIS Benchmarks). Depending on the document type, those requirements might be called differently. In internal security policies, security requirements are referred to as clauses, in security standards they often are called controls, and in public guidelines they are often coined recommendations. Hence, when we use the terms \textit{clauses}, \textit{controls}, or \textit{recommendations}, we refer to security requirements in the context of the respective document type formalizing those requirements.

\subsection{Automated Detection of Security-Relevant Backlog Items}

Another branch of prior work focuses on automatically identifying security-relevant backlog items.
Wu et al.~\cite{wu2022data-quality} showed that widely used datasets contain substantial label noise and released cleaned datasets that have since become a benchmark for state-of-the-art approaches~\cite{alqahtani2024fasttext, franca2025gpts, soltaniani2026evaluating-llms}.
Subsequent work has improved backlog item-based security-relevance identification on those benchmarks using lightweight neural text models~\cite{alqahtani2024fasttext}, open-source GPT-based approaches~\cite{franca2025gpts}, and fine-tuned or prompted LLMs~\cite{soltaniani2026evaluating-llms}.

Relevant to our contributions are other detection approaches focusing on the privacy domain.
Sangaroonsilp et al.~\cite{sangaroonsilp2023privacy-issue-reports} classify privacy requirements in backlog items by treating issue trackers as the central agile artifact and aligning backlog items with privacy-requirement categories to support compliance-oriented development.

A second closely related line of work makes regulations actionable for engineers.
Ayala-Rivera and Pasquale's GuideMe approach~\cite{ayala-rivera2018grace-period} supports the operationalization of GDPR obligations into solution requirements and privacy controls, and the SoCo approach~\cite{ayala-rivera2024soco} incorporates suitable privacy and security requirements into software design models.

These approaches share the motivation of C3 but operate at the regulation-to-requirement or design-model level, rather than on the backlog level.

Finally, software traceability research helps explain why this problem remains open in practice.
Ruiz et al.~\cite{ruiz2023traceability} report that practitioners still perceive traceability as valuable but costly, with substantial manual effort and limited automated support.
This strengthens the motivation for tooling that links the artifacts developers already use to the security requirements they are expected to satisfy.
\section{Methodology}

\begin{figure*}
    \centering
    \includegraphics[width=0.80\linewidth]{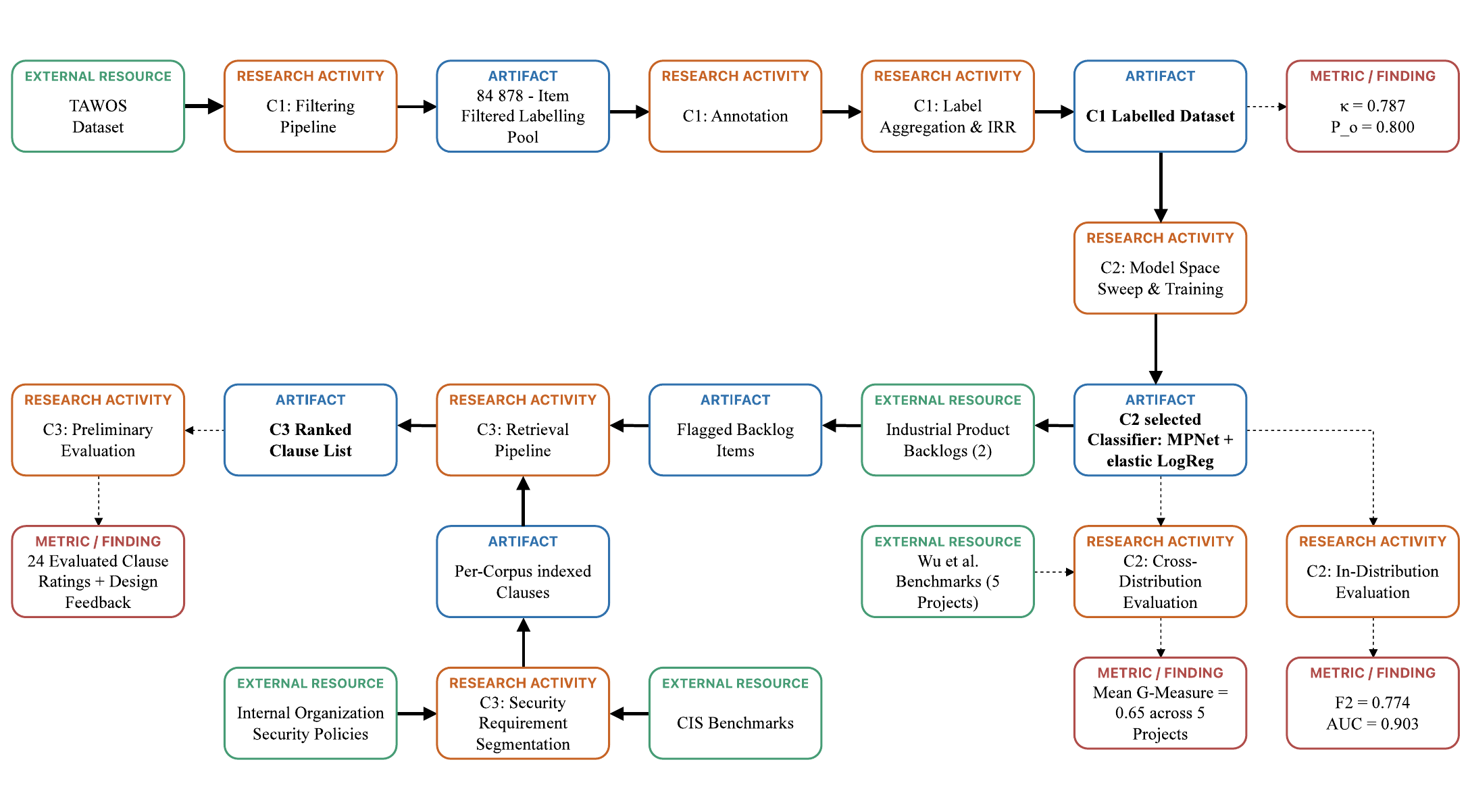}
    \caption{Overview of Contributions and Research Activities as a Chained Workflow}
    \Description[Flowchart chaining contributions C1, C2 and C3, from the TAWOS dataset through classifier training to security clause retrieval and evaluation]{A flowchart showing how the three contributions chain together into a single workflow. Nodes are typed and colour-coded: green nodes are external resources, orange nodes are research activities, blue nodes are artifacts produced by this work, and red nodes are metrics or findings. Solid arrows carry the main workflow; dashed arrows lead to evaluation results. The C1 strand runs left to right along the top. The external TAWOS dataset feeds a C1 filtering pipeline, which produces an 84,878-item filtered labelling pool. That pool is annotated in the C1 annotation activity, and a label aggregation and inter-rater reliability activity then produces the C1 labelled dataset. A dashed arrow from the dataset reports Fleiss' kappa of 0.787 and observed agreement of 0.800. The C2 strand descends from the C1 labelled dataset: a model space sweep and training activity yields the selected classifier, an MPNet encoder with an elastic-net logistic regression head. Two dashed evaluation branches leave that classifier. The first leads to a cross-distribution evaluation, which additionally takes the external Wu et al. benchmarks covering five projects as input and reports a mean G-measure of 0.65 across those five projects. The second leads to an in-distribution evaluation, reporting an F2 score of 0.774 and an AUC of 0.903. The C3 strand runs right to left across the lower half. The selected classifier is applied to two industrial product backlogs, producing flagged backlog items, which enter the C3 retrieval pipeline. That pipeline is fed from below by a second path: internal organisation security policies and the CIS Benchmarks, both external resources, are processed by a security requirement segmentation activity into per-corpus indexed clauses, which in turn feed the retrieval pipeline. The pipeline outputs a C3 ranked clause list, which a dashed arrow passes to the C3 preliminary evaluation, producing 24 evaluated clause ratings together with design feedback.}
    \label{fig:methodology-overview}
\end{figure*}

\Cref{fig:methodology-overview} summarizes the three contributions as a chained workflow: C1 produces the expert-labeled dataset on which C2 is trained.
C2 flags security-relevant items from a backlog, and C3 retrieves applicable security requirements for each flagged backlog item.
The subsections below describe the artifacts, activities, and evaluation protocol of each contribution.

\subsection*{C1: Annotated Dataset of Security-Relevant Backlog Items}

Contemporary approaches~\cite{alqahtani2024fasttext,franca2025gpts,soltaniani2026evaluating-llms} evaluate against the Wu et al.\ benchmark~\cite{wu2022data-quality}, whose security labels derive from the Jira \texttt{security} field in the dataset rather than from expert annotation.
C1 contributes a smaller but independently expert-annotated dataset.

We draw from TAWOS~\cite{tawosi2022tawos}, a public archive of over $500{,}000$ backlog items from $44$ open-source agile projects, selected for its scale, diversity, and redistributability.

\textbf{Source Corpus.}
From this pool we selected two Atlassian projects, \emph{Jira Server} and \emph{Confluence Server}, whose items describe enterprise software tooling and whose language resembles the industrial backlogs targeted by C3.
A filtering pipeline applies title-level and full-text exact deduplication and removes near-duplicates via MinHash LSH ($128$ permutations, $5$-word shingles) at a Jaccard threshold of $0.8$, yielding an $84{,}878$-item labeling pool.

\textbf{Labeling protocol.}
Nine practitioners from a large enterprise were recruited on a voluntary and uncompensated basis.
A web application distributes items for asynchronous, independent labeling over a two-week window, and a kick-off session introduced the task, boundary cases, and the tool.
For each item, labelers record a binary \texttt{security-relevant} judgment after reading the title and description.
One of the initial hypotheses was that security-requirement categories might improve the ultimately retrieved security requirements.
For that purpose, practitioners additionally assigned a Firesmith security-requirement category~\cite{firesmith2004specifying} to each backlog item.
As this hypothesis did not hold, we only store the Firesmith tags as metadata in the released artifact but did not use them further, e.g., as training labels.
Reliability and agreement checks are not performed on this metadata.

\textbf{Label aggregation and reliability.}
Items require at least two independent assessments before release. As the labeling process was distributed asynchronously across a pool of nine practitioners, not every practitioner rated every backlog item. Reliability statistics are therefore computed over the available overlapping ratings in the released dataset.
The published label is the majority vote of labelers' binary judgments.
Following Artstein and Poesio~\cite{artstein2008irr}, we assess inter-rater reliability (IRR) with a generalized multi-rater coefficient rather than averaged pairwise coefficients, and additionally report observed agreement for transparency.
Accordingly, for C1 we report Fleiss' $\kappa$, which quantifies  agreement beyond chance across all raters simultaneously, and raw observed agreement $P_o$, which provides an interpretable baseline of direct rater consensus.

\subsection*{C2: Security-Relevance Detection Tool}\label{sec:methodology-c2}

C2 is a lightweight binary classifier that flags potentially security-relevant backlog items as a first-stage filter for the downstream backlog enrichment system.
Because it is more costly for an engineer to miss relevant security requirements for a backlog item than to discard irrelevant security requirements attached to it, we optimize C2 for recall-weighted performance and use F2 as the primary evaluation criterion, but do not tune the decision threshold as the validation split is too small for reliable threshold calibration.

The input is the concatenated title and description of a backlog item.
The output is a binary security-relevance prediction, with C1 as the gold standard.

\textbf{Architecture.}
We pair a frozen sentence-transformer encoder with a lightweight classification head, deliberately avoiding full fine-tuning in order to retain cross-domain generalizability.

The reported C2 model space comprises one TF-IDF baseline and eight sentence-encoder variants.
The sentence encoders are MiniLM-L6, MiniLM-L12, MPNet, DistilRoBERTa-v1, and RoBERTa-Large, all used via the Sentence Transformers library~\cite{reimers2019sbert}.
All five encoders are evaluated with an elastic-net logistic-regression head.
In addition, DistilRoBERTa-v1 and MPNet are also evaluated with a small Multilayer Perceptron (MLP) head, and DistilRoBERTa-v1 is additionally evaluated with an $\ell_2$-regularized logistic-regression head.

\textbf{Training protocol.}
The protocol follows established practices for recall-oriented binary classification on imbalanced software engineering datasets~\cite{tantithamthavorn2020rebalancing}.
We apply a stratified hold-out split of C1 with class-balanced sample weighting during training.
In the used dataset, security-relevant backlog items have a prevalence of $27.8\%$.
Regularization strength (for logistic regression) and MLP width and depth are tuned via grid search over 5-fold stratified cross-validation on the training portion.
All models are evaluated at a fixed threshold of $0.5$.

For deployment selection, we treat the elastic-net and MLP variants as the transfer-oriented candidate set.
The choice of an elastic-net head over $\ell_2$-regularized logistic regression and MLP is motivated by generalization risk on a small dataset: elastic-net's combined $\ell_1$ and $\ell_2$ penalty promotes sparsity and resists overfitting to the two-project scope of C1, whereas $\ell_2$-only regularization and unconstrained MLP heads are more susceptible to absorbing project-specific lexical patterns~\cite{zou2005elastic-net}.
We therefore report $\ell_2$ results as an in-distribution reference, but do not treat the $\ell_2$ variant as deployment candidate.

\textbf{Evaluation.}
In-distribution performance is reported on the held-out C1 test set ($n=58$ and prevalence of the \texttt{security-relevant} class consistent with a stratified split over C1).
The primary metric is F2 (the $F_\beta$ score at $\beta = 2$, which weights recall twice as heavily as precision).
Secondary metrics are precision, recall, AUC, and per-model confusion matrices.

To test whether the learned notion of \texttt{security-relevant} transfers across projects, we evaluate the reported C2 variants zero-shot on the five Wu et al.~\cite{wu2022data-quality} datasets (Chromium, Ambari, Camel, Derby, Wicket) and compare against four families of published baselines: Wu et al.~\cite{wu2022data-quality}: (FARSEC variants and classical text classifiers), Soltaniani et al.~\cite{soltaniani2026evaluating-llms} (prompted proprietary LLMs and fine-tuned LLMs), Alqahtani~\cite{alqahtani2024fasttext} (fastText), and França et al.~\cite{franca2025gpts} (open-source GPTs and classical ML).
The held-out C1 F2 ranking is used separately to choose the variant integrated into C3.

The primary cross-distribution metric is G-measure (harmonic mean of recall and $1{-}\mathrm{FPR}$), which is insensitive to the strong class imbalance present in both C1 and the Wu et al. datasets ($27.8\%$ vs.\ $1.9$--$17.9\%$).
All four baseline families report or allow recomputation of G-measure, making it the single comparable metric across literature.
We additionally report the per-project standard deviation $\sigma$ of G-measure as an indicator of cross-project stability, with the caveat that França et al.~\cite{franca2025gpts} do not evaluate on Chromium; their $\sigma$ is therefore computed over four projects rather than five and is not directly comparable to the other baseline families on equal footing.

\subsection*{C3: Backlog Item Enrichment}\label{sec:methodology-c3}

Given a security-relevant backlog item, C3 surfaces the subset of security requirements from relevant security requirements documents.
This should help engineers and security experts to address security and compliance proactively during backlog refinement and planning, rather than through reactive checks later in the development lifecycle, directly addressing the lead-time and workload burdens reported by Moyón et al.~\cite{moyon2024challenges}.

We frame this contribution as a document-grounded information retrieval problem: given the title and description of a backlog item, retrieve and rank the most relevant requirements from a corpus of segmented security documents.

The backlog enrichment pipeline intended for C3 is developed for on-premises deployment and operation. It is designed as a four-stage retrieval-augmented generation (RAG)~\cite{lewis2020rag} architecture applied per backlog item.
\begin{enumerate}
    \item \emph{Requirement extraction.} An open-weights LLM decomposes the backlog-item text into one or more requirements, so multi-concern items do not dilute their retrieval signal across competing topics.
    \item \emph{Bi-encoder recall.} A sentence-transformer bi-encoder~\cite{reimers2019sbert} embeds each extracted requirement and retrieves the top-$k$ candidate security requirements by exact cosine similarity over precomputed embeddings of all segmented security requirements.
    \item \emph{Cross-encoder reranking.} A sentence-pair cross-encoder~\cite{nogueira2019bert} rescores the shortlist with direct query--security requirement attention, sharpening the ranking over the dense baseline.
    \item \emph{LLM semantic filter.} An open-weights LLM reviews the reranked shortlist and retains only security requirements it judges semantically applicable to the requirement, producing the final ranked list. The LLM acts as a precision filter, not a reranker: its role is to retain or discard individual security requirements, grounded in the bi-encoder shortlist so identifiers cannot be hallucinated.
\end{enumerate}

The architecture was reached iteratively.
\emph{Version~1 (Dense IR only)} uses the bi-encoder and cross-encoder as a standard two-stage retrieval baseline; it produces stable shortlists but admits topically adjacent false positives where lexical or vector similarity is high but the security requirement does not impose the same requirement.
\emph{Version~2} adds the LLM semantic filter to prune the reranked shortlist with a brief per-security requirement justification.
\emph{Version~3} adds requirement extraction as a preprocessing step, introduced after we observed that multi-concern backlog items (e.g., ``add OAuth2 login \emph{and} rate-limit the public API'') systematically retrieved security requirements relevant to only one of their concerns. Decomposing into atomic requirements before retrieval recovers the missed concerns.

Both the extractor and the filter use open-weights LLMs so the pipeline can be deployed on-premises, satisfying a hard privacy requirement for the industrial deployment context.
Additionally, pinning a specific open model version prevents silent behavior drifts known from commercial models~\cite{baltes2026guidelinesempiricalstudiessoftware, angermeirReproducibility}.
For both requirement extraction and LLM-based clause scoring in C3, we used \texttt{alibayram/Qwen3-30B-A3B-Instruct-2507}, identified by the Ollama model ID \texttt{408ee351bdc6} (architecture \texttt{qwen3moe}, $30.5$B parameters, \texttt{Q4\_K\_M} quantization).

\textbf{Security Requirement Segmentation.}
Segmentation is a one-time, document-specific preprocessing step that converts each security document into a flat list of atomic, self-contained security requirements suitable for retrieval.
We provide a segmentation backend for CIS Benchmarks~\cite{cis2024benchmarks} exported as PDF files, and the retrieval pipeline itself is segmentation-agnostic.
The evaluation reported in this paper uses the internal-policy and CIS Benchmark backends.

\subsubsection*{Preliminary Evaluation Protocol}\label{sec:methodology-c3-eval}

We report a preliminary evaluation on the performance and workflow fit potential of C3.
Insights gained on this analysis are then used to further develop the tool and ensure seamless and efficient integration into continuous software engineering workflows.

The preliminary evaluation combines an artifact-level analysis of the pipeline's retrieval output with two semi-structured practitioner interviews conducted on industrial backlogs.
The interviews contextualize the retrieval results and help identify design requirements that may be necessary for broader adoption.
The units of analysis are retrieved clauses and controls in relation to their source backlog items.

\textbf{Evaluation criteria.}
We define success of the preliminary evaluation along three dimensions: \emph{relevance}, whether a retrieved security requirement applies to the backlog item, captured by the practitioner rating; \emph{actionability}, whether the requirement is concrete enough to change the engineer's next action on the item, captured by the upper rating anchor and the free-text commentary, and \emph{workflow fit}, whether practitioners would intend to consume the output within their existing tools, captured by the closing discussion questions.
Redundancy and coverage gaps are recorded through the free-text commentary rather than quantified.
The ratings serve as qualitative indicators for future design iteration.
Quantifying retrieval precision with more raters, items, and backlogs is deferred to future work.

\textbf{Source corpora.}
We evaluate on two complementary corpora: internal security policies as the primary use case, and public CIS Benchmarks (nginx, docker-ce configurations) as a probe for generalization to concrete best-practice guidelines that any reader can reproduce against.
Using both corpora in the same interview session, on the same backlog items, allows a direct within-session comparison of retrieval quality across security document types without modifying the retrieval pipeline.
Only the security requirement segmenter backend has to be swapped.

\textbf{Item selection.}
Items for each session were drawn randomly from the set of backlog items that the selected C2 classifier flagged at confidence scores ranging from $0.55$ to $0.93$, spanning high-confidence and low-confidence regions of the classifier's output.
This selection ensures that evaluated items are representative of what the pipeline would surface in normal operation.

\textbf{Interview sessions.}
Two $30$-minute interview sessions were conducted with practitioners from the product team owning the evaluated backlogs.
In preparation, the deployed pipeline was run over each product's backlog using both corpora, the internal policies and CIS Benchmarks.
Each session presents the resulting shortlists and collects two structured signals per security requirement: a $0$--$5$ relevance rating ($0 =$ not relevant, $5 =$ directly actionable) and free-text commentary capturing the reasoning.
Item-level familiarity is recorded as descriptive context, but is not used to weight findings; the sessions target the scenario in which the tool is intended to be used, where the consumer of a security requirement might or might not be the author of the corresponding backlog item, as they might be an engineer or a security expert.

Three open discussion questions close each session: (D1)~workflow fit and actionability, (D2)~missing security requirements or concerns, and (D3)~open feedback, including trust and integration preferences.

A web application supported the sessions by loading the product backlog, retrieving relevant security requirements, and presenting them.

\section{Results}
We report the results contribution by contribution: the dataset characteristics and inter-rater reliability of C1, the in-distribution and cross-distribution behavior of the C2 classifier, and the retrieval quality and practitioner design feedback gathered for C3.

\subsection*{C1 Results}
The C1 release contains $288$ backlog items ($80$ \texttt{security-relevant}, $27.8\%$), drawn from an $84{,}878$-item filtered TAWOS pool and annotated by a pool of nine security experts.
Each released item received at least two independent assessments, resulting in 972 binary security-relevance judgments in total. 
Inter-rater reliability on the released items is substantial: Fleiss' $\kappa = 0.787$, and raw observed agreement $P_o = 0.800$, which confirms a reliable but non-trivial task.

\begin{figure}
    \centering
    \includegraphics[width=0.8\linewidth]{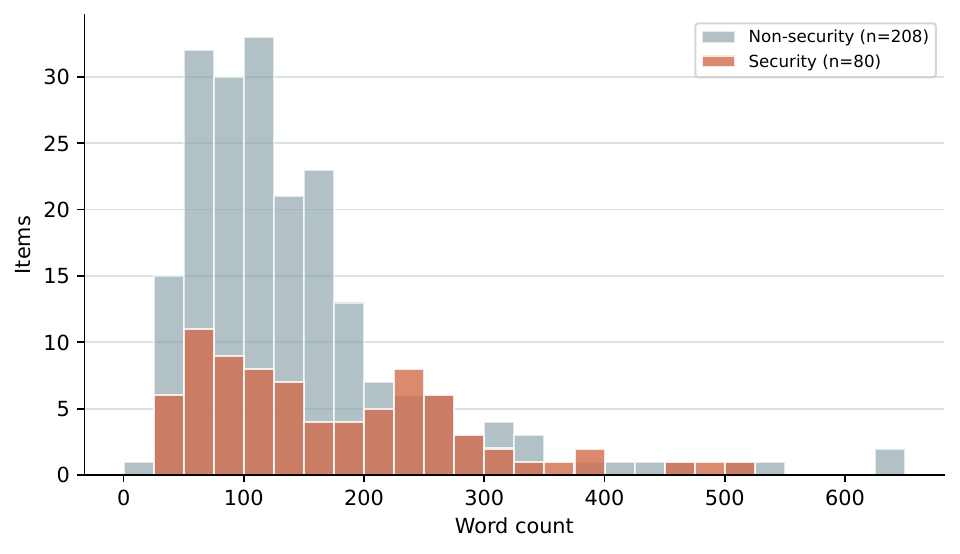}
    \caption[Frequency of Labels based on Text Length on C1]{Frequency of Labels based on Text Length on C1, in Groups of 25 Words.}
    \Description[Stacked histogram of C1 backlog item word counts, split by security and non-security labels, with both classes concentrated below 200 words and a long right tail]{A stacked histogram of the 288 C1 backlog items, binned by description length in groups of 25 words. The horizontal axis is word count, running from 0 to about 650. The vertical axis is the number of items, running from 0 to about 33. Each bar is stacked from two series given in the legend: non-security items, 208 in total, drawn in pale blue-grey, and security-relevant items, 80 in total, drawn in terracotta. The distribution is strongly right-skewed. Almost all mass falls between 25 and 300 words, with the tallest bars, around 30 to 33 items each, in the bins spanning roughly 50 to 125 words. Counts fall away steadily above 150 words. Below about 200 words the bars are dominated by the pale non-security series; from roughly 200 to 300 words the terracotta security series makes up most or all of each bar, so security-relevant items are proportionally more common among longer descriptions. Beyond 300 words the data thin to isolated bars of one or two items, extending to a final non-security bar near 650 words.}
    \label{fig:c1-text-length}
\end{figure}

Of the $288$ items, $249$ are unanimous ($55$ \texttt{security-relevant}, $194$ \texttt{non-security-relevant}) and $39$ ($13.5\%$) show at least one dissenting assessor. Disagreement items are retained with the majority-vote label.
As \Cref{fig:c1-text-length} highlights, text descriptions are long enough to support text-classification methods: median $124$ words (range $21$--$748$); security-relevant backlog items are meaningfully longer (median $145$) than non-security-relevant backlog items (median $119$), suggesting implicit security concerns surface more often in backlog items with richer contextual descriptions.

\begin{table}[t]
    \centering
    \caption[C1 Dataset Characteristics Relative to the Wu et al.\ Benchmark Family]{C1 Dataset Characteristics Relative to the Wu et al.\ Benchmark Family. C1 is Smaller but Provides Multi-Rater Expert Annotation with Reported IRR.}
    \Description[Table comparing C1 with the five Wu et al. datasets by item count, security-relevant count and prevalence]{A table with five columns: dataset, source, number of items, number of security-relevant items, and prevalence. The first row, highlighted, is C1 (ours), drawn from Atlassian Jira and Confluence: 288 items, 80 security-relevant, 27.8 percent prevalence. The remaining five rows are the Wu et al. benchmarks. Chromium, from Chromium bugs, has 41,940 items, 808 security-relevant, 1.9 percent. Ambari, Camel and Wicket, all from the Apache JIRA, each have 1,000 items, with 56, 74 and 47 security-relevant items and prevalences of 5.6, 7.4 and 4.7 percent respectively. Derby, also Apache JIRA, has 1,000 items, 179 security-relevant, 17.9 percent. C1 is by far the smallest dataset but has the highest prevalence of security-relevant items.}
    \label{fig:c1-wu-comparison}
    % Auto-generated from c1_dataset_analysis.ipynb
% Requires: \usepackage[table]{xcolor}
% Requires: \usepackage{graphicx}
\begingroup
\setlength{\tabcolsep}{4pt}
\renewcommand{\arraystretch}{1.08}
\definecolor{litC1Ours}{HTML}{E8F4E8}
\resizebox{\linewidth}{!}{%
\begin{tabular}{llrrc}
\hline
Dataset & Source & Items & Sec. & Prevalence \\
\hline
\cellcolor{litC1Ours}\textbf{C1 (Ours)} & \cellcolor{litC1Ours}\textbf{Atlassian Jira / Confluence} & \cellcolor{litC1Ours}\textbf{288} & \cellcolor{litC1Ours}\textbf{80} & \cellcolor{litC1Ours}\textbf{27.8\%} \\
Wu: Chromium & Chromium bugs & 41940 & 808 & 1.9\% \\
Wu: Ambari & Apache JIRA & 1000 & 56 & 5.6\% \\
Wu: Camel & Apache JIRA & 1000 & 74 & 7.4\% \\
Wu: Derby & Apache JIRA & 1000 & 179 & 17.9\% \\
Wu: Wicket & Apache JIRA & 1000 & 47 & 4.7\% \\
\hline
\end{tabular}%
}
\endgroup

\end{table}

Relative to the Wu et al.\ family of benchmarks ($1{,}000$--$41{,}940$ items, $1.9$--$17.9\%$ prevalence, community-assigned labels), C1 is smaller, but to the best of our knowledge, the only publicly released backlog-security dataset with independent multi-rater expert annotation and reported IRR, positioning it as a high-precision training dataset that complements Wu's larger-scale datasets.
An overview of the items and prevalence of the \texttt{security-relevant} label across C1 and the datasets from Wu et al.~\cite{wu2022data-quality} is available at \Cref{fig:c1-wu-comparison}.

\subsection*{C2 Results}

\begin{figure}
    \centering
    \includegraphics[width=0.75\linewidth]{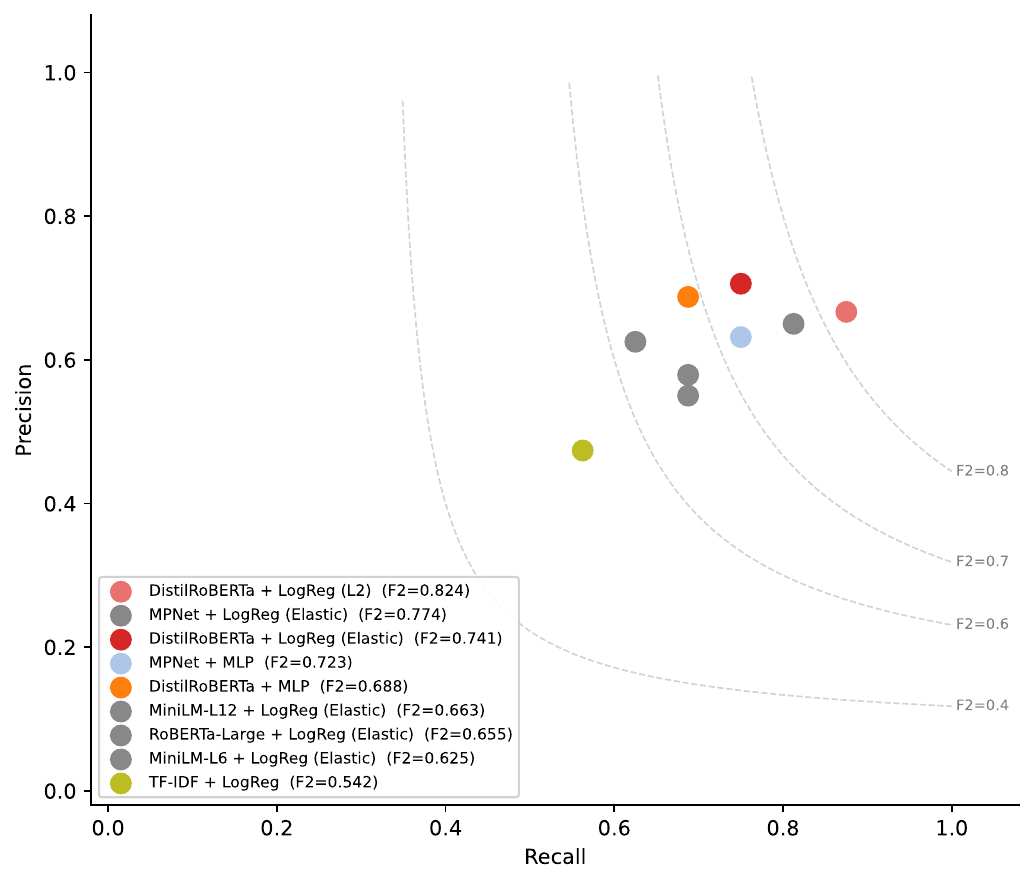}
    \caption{In-Distribution (C1 Held-Out) Precision/Recall of All C2 Variants at the Fixed Operating Threshold of $0.5$.}
    \Description[Precision-recall scatter plot of nine C2 classifier variants at a fixed threshold of 0.5, overlaid with iso-F2 contour curves]{A scatter plot comparing nine classifier variants on the held-out C1 test split at a fixed operating threshold of 0.5. The horizontal axis is recall and the vertical axis is precision, both from 0.0 to 1.0. Four dashed grey contour curves sweep across the plot marking constant F2 scores, labelled at the right edge as F2 equals 0.4, 0.6, 0.7 and 0.8; curves further toward the upper right indicate better performance. Each variant is one coloured dot, identified in the legend together with its F2 score, in descending order: DistilRoBERTa with L2 logistic regression, F2 of 0.824; MPNet with elastic-net logistic regression, F2 of 0.774; DistilRoBERTa with elastic-net logistic regression, F2 of 0.741; MPNet with an MLP head, F2 of 0.723; DistilRoBERTa with an MLP head, F2 of 0.688; MiniLM-L12 with elastic-net logistic regression, F2 of 0.663; RoBERTa-Large with elastic-net logistic regression, F2 of 0.655; MiniLM-L6 with elastic-net logistic regression, F2 of 0.625; and a TF-IDF with logistic regression baseline, F2 of 0.542. All nine points fall in a loose cluster spanning roughly 0.55 to 0.87 recall and 0.47 to 0.71 precision. The two best variants sit furthest to the right, trading a little precision for markedly higher recall, while the TF-IDF baseline sits alone at the lower left, below every neural variant on both axes.}
    \label{fig:c2-in-dist}
\end{figure}

The best in-distribution F2 overall is achieved by DistilRoBERTa + LogReg (L2).
However, as determined in \Cref{sec:methodology-c2}, $\ell_2$ variants are treated as in-distribution references rather than deployment candidates because they are more exposed to project-specific overfitting on the two-project scope of C1.
Within the transfer-oriented deployment candidate set, \textbf{MPNet + elastic-net logistic regression} achieves the highest held-out F2 and is therefore selected at threshold $0.5$ for integration into C3.
On the held-out C1 test set ($n=58$) it reaches F2 $=0.774$, precision $0.650$, recall $0.812$, and ROC-AUC $0.903$. This poses an improvement over the TF-IDF+LogReg baseline (F2 $=0.542$, AUC $0.750$) and comparable to the other sentence-embedder-based variants explored.
Across all $8$ sentence-encoder variants, F2 clusters between $0.625$ and $0.824$ with AUC between $0.839$ and $0.939$, suggesting that the task is well-posed and the signal is recoverable with any competitive text representation (see \Cref{fig:c2-in-dist}).

\subsubsection*{Cross-Distribution Generalizability}

\Cref{fig:c2-wu-summary} summarizes mean G-measure and cross-project standard deviation across two of our models and the literature baselines.

Applied to the Wu et al. benchmarks~\cite{wu2022data-quality}, MPNet + LogReg (Elastic) achieves mean G-measure $=0.649$ (per-dataset: Ambari $0.716$, Camel $0.58$, Chromium $0.80$, Derby $0.5$, and Wicket $0.64$), with mean per-project recall of $0.771$ (ranging from $0.66$ to $0.91$), despite a prevalence shift from $27.8\%$ (C1) to $1.9\%$--$17.9\%$ (Wu et al. datasets~\cite{wu2022data-quality}). 

Against the published baselines on the same benchmarks, MPNet + LogReg (Elastic)'s mean G-measure ($0.65$) is on par with the strongest classical ML baseline from Wu et al.~\cite{wu2022data-quality} (FARSEC, mean $0.64$), with a mixed per-project picture: it outperforms all of Wu et al.'s classical ML baselines (FARSEC + RF/NB/LR/KNN/MLP) on Ambari and Camel, is competitive on Chromium and Wicket without exceeding the best FARSEC variants, and trails Wu et al.'s baselines on Derby.
Against França et al.~\cite{franca2025gpts}, MPNet + LogReg (Elastic) outperforms every open-source GPT zero-shot and classical ML baseline on Ambari and Wicket, ties with França's strongest classical baseline (RF) on Camel, and trails the same baseline on Derby.
It is competitive with the best fine-tuned small LLMs (BERT, DistilBERT) from Soltaniani et al.~\cite{soltaniani2026evaluating-llms} while being a frozen-encoder with a linear head (no LLM fine-tuning, orders of magnitude cheaper).
MPNet + LogReg (Elastic) does not outperform the best-performing Gemini 2.5 configuration from Soltaniani et al.~\cite{soltaniani2026evaluating-llms}.
Alqahtani's~\cite{alqahtani2024fasttext} \texttt{fastText} in-distribution $10$-fold CV numbers are also higher, though that comparison is not cross-distribution and therefore not directly informative.

\begin{table}[t]
    \centering
    \caption{Per-Project G-Measure Heatmap Across All Evaluated Models on the Five Wu et al.\ Benchmarks.}
    \label{fig:c2-wu-summary}
    % Auto-generated from generalizability_evaluation.ipynb
% Requires: \usepackage[table]{xcolor}
% Requires: \usepackage{graphicx}
\begingroup
\setlength{\tabcolsep}{4pt}
\renewcommand{\arraystretch}{1.08}
% - = dataset not evaluated in that paper | Mean and sigma over available datasets
\definecolor{litWuEtAl}{HTML}{DDE8F2}
\definecolor{litWuEtAlSummary}{HTML}{C7D1DA}
\definecolor{litSoltanianiEtAl}{HTML}{E8DDF2}
\definecolor{litSoltanianiEtAlSummary}{HTML}{D1C7DA}
\definecolor{litAlqahtani}{HTML}{F2E8DD}
\definecolor{litAlqahtaniSummary}{HTML}{DAD1C7}
\definecolor{litFranaEtAl}{HTML}{DDF2E8}
\definecolor{litFranaEtAlSummary}{HTML}{C7DAD1}
\definecolor{litOurs}{HTML}{A8DBA8}
\definecolor{litOursSummary}{HTML}{97C597}
\resizebox{\linewidth}{!}{%
\begin{tabular}{lllccccc!{\color[HTML]{444444}\vrule width 1.5pt}cc}
\hline
Paper & Group & Method & Chromium & Ambari & Camel & Derby & Wicket & Mean & $\sigma$ \\
\hline
\cellcolor{litWuEtAl}Wu et al. & \cellcolor{litWuEtAl}FARSEC baselines & \cellcolor{litWuEtAl}Farsec & \cellcolor{litWuEtAl}0.79 & \cellcolor{litWuEtAl}0.65 & \cellcolor{litWuEtAl}0.48 & \cellcolor{litWuEtAl}0.60 & \cellcolor{litWuEtAl}0.67 & \cellcolor{litWuEtAlSummary}0.64 & \cellcolor{litWuEtAlSummary}0.10 \\
\cellcolor{litWuEtAl}Wu et al. & \cellcolor{litWuEtAl}FARSEC baselines & \cellcolor{litWuEtAl}Farsec+TunedLearner & \cellcolor{litWuEtAl}0.80 & \cellcolor{litWuEtAl}0.70 & \cellcolor{litWuEtAl}0.38 & \cellcolor{litWuEtAl}0.62 & \cellcolor{litWuEtAl}0.67 & \cellcolor{litWuEtAlSummary}0.63 & \cellcolor{litWuEtAlSummary}0.14 \\
\cellcolor{litWuEtAl}Wu et al. & \cellcolor{litWuEtAl}FARSEC baselines & \cellcolor{litWuEtAl}Farsec+TunedSMOTE & \cellcolor{litWuEtAl}0.86 & \cellcolor{litWuEtAl}0.60 & \cellcolor{litWuEtAl}0.47 & \cellcolor{litWuEtAl}0.67 & \cellcolor{litWuEtAl}0.55 & \cellcolor{litWuEtAlSummary}0.63 & \cellcolor{litWuEtAlSummary}0.13 \\
\cellcolor{litWuEtAl}Wu et al. & \cellcolor{litWuEtAl}Text classifiers & \cellcolor{litWuEtAl}RF & \cellcolor{litWuEtAl}0.84 & \cellcolor{litWuEtAl}0.40 & \cellcolor{litWuEtAl}0.54 & \cellcolor{litWuEtAl}0.63 & \cellcolor{litWuEtAl}0.65 & \cellcolor{litWuEtAlSummary}0.61 & \cellcolor{litWuEtAlSummary}0.14 \\
\cellcolor{litWuEtAl}Wu et al. & \cellcolor{litWuEtAl}Text classifiers & \cellcolor{litWuEtAl}NB & \cellcolor{litWuEtAl}0.79 & \cellcolor{litWuEtAl}0.60 & \cellcolor{litWuEtAl}0.46 & \cellcolor{litWuEtAl}0.66 & \cellcolor{litWuEtAl}0.46 & \cellcolor{litWuEtAlSummary}0.59 & \cellcolor{litWuEtAlSummary}0.13 \\
\cellcolor{litWuEtAl}Wu et al. & \cellcolor{litWuEtAl}Text classifiers & \cellcolor{litWuEtAl}MLP & \cellcolor{litWuEtAl}0.66 & \cellcolor{litWuEtAl}0.54 & \cellcolor{litWuEtAl}0.26 & \cellcolor{litWuEtAl}0.55 & \cellcolor{litWuEtAl}0.36 & \cellcolor{litWuEtAlSummary}0.47 & \cellcolor{litWuEtAlSummary}0.14 \\
\cellcolor{litWuEtAl}Wu et al. & \cellcolor{litWuEtAl}Text classifiers & \cellcolor{litWuEtAl}LR & \cellcolor{litWuEtAl}0.73 & \cellcolor{litWuEtAl}0.40 & \cellcolor{litWuEtAl}0.26 & \cellcolor{litWuEtAl}0.55 & \cellcolor{litWuEtAl}0.23 & \cellcolor{litWuEtAlSummary}0.43 & \cellcolor{litWuEtAlSummary}0.19 \\
\cellcolor{litWuEtAl}Wu et al. & \cellcolor{litWuEtAl}Text classifiers & \cellcolor{litWuEtAl}KNN & \cellcolor{litWuEtAl}0.73 & \cellcolor{litWuEtAl}0.40 & \cellcolor{litWuEtAl}0.26 & \cellcolor{litWuEtAl}0.55 & \cellcolor{litWuEtAl}0.23 & \cellcolor{litWuEtAlSummary}0.43 & \cellcolor{litWuEtAlSummary}0.19 \\
\cellcolor{litSoltanianiEtAl}Soltaniani et al. & \cellcolor{litSoltanianiEtAl}Prompted LLMs & \cellcolor{litSoltanianiEtAl}GPT-4.1 & \cellcolor{litSoltanianiEtAl}0.48 & \cellcolor{litSoltanianiEtAl}0.40 & \cellcolor{litSoltanianiEtAl}0.23 & \cellcolor{litSoltanianiEtAl}0.41 & \cellcolor{litSoltanianiEtAl}0.29 & \cellcolor{litSoltanianiEtAlSummary}0.36 & \cellcolor{litSoltanianiEtAlSummary}0.09 \\
\cellcolor{litSoltanianiEtAl}Soltaniani et al. & \cellcolor{litSoltanianiEtAl}Prompted LLMs & \cellcolor{litSoltanianiEtAl}Gemini-2.5 & \cellcolor{litSoltanianiEtAl}0.86 & \cellcolor{litSoltanianiEtAl}0.74 & \cellcolor{litSoltanianiEtAl}0.73 & \cellcolor{litSoltanianiEtAl}0.73 & \cellcolor{litSoltanianiEtAl}0.81 & \cellcolor{litSoltanianiEtAlSummary}0.77 & \cellcolor{litSoltanianiEtAlSummary}0.05 \\
\cellcolor{litSoltanianiEtAl}Soltaniani et al. & \cellcolor{litSoltanianiEtAl}Fine-tuned LLMs & \cellcolor{litSoltanianiEtAl}BERT & \cellcolor{litSoltanianiEtAl}0.83 & \cellcolor{litSoltanianiEtAl}0.40 & \cellcolor{litSoltanianiEtAl}0.33 & \cellcolor{litSoltanianiEtAl}0.57 & \cellcolor{litSoltanianiEtAl}0.23 & \cellcolor{litSoltanianiEtAlSummary}0.47 & \cellcolor{litSoltanianiEtAlSummary}0.21 \\
\cellcolor{litSoltanianiEtAl}Soltaniani et al. & \cellcolor{litSoltanianiEtAl}Fine-tuned LLMs & \cellcolor{litSoltanianiEtAl}DistilBERT & \cellcolor{litSoltanianiEtAl}0.85 & \cellcolor{litSoltanianiEtAl}0.32 & \cellcolor{litSoltanianiEtAl}0.33 & \cellcolor{litSoltanianiEtAl}0.56 & \cellcolor{litSoltanianiEtAl}0.47 & \cellcolor{litSoltanianiEtAlSummary}0.51 & \cellcolor{litSoltanianiEtAlSummary}0.19 \\
\cellcolor{litSoltanianiEtAl}Soltaniani et al. & \cellcolor{litSoltanianiEtAl}Fine-tuned LLMs & \cellcolor{litSoltanianiEtAl}DistilGPT2 & \cellcolor{litSoltanianiEtAl}0.88 & \cellcolor{litSoltanianiEtAl}0.12 & \cellcolor{litSoltanianiEtAl}0.04 & \cellcolor{litSoltanianiEtAl}0.18 & \cellcolor{litSoltanianiEtAl}0.04 & \cellcolor{litSoltanianiEtAlSummary}0.25 & \cellcolor{litSoltanianiEtAlSummary}0.32 \\
\cellcolor{litSoltanianiEtAl}Soltaniani et al. & \cellcolor{litSoltanianiEtAl}Fine-tuned LLMs & \cellcolor{litSoltanianiEtAl}Qwen2.5B & \cellcolor{litSoltanianiEtAl}0.87 & \cellcolor{litSoltanianiEtAl}0.22 & \cellcolor{litSoltanianiEtAl}0.26 & \cellcolor{litSoltanianiEtAl}0.43 & \cellcolor{litSoltanianiEtAl}0.30 & \cellcolor{litSoltanianiEtAlSummary}0.42 & \cellcolor{litSoltanianiEtAlSummary}0.24 \\
\cellcolor{litAlqahtani}Alqahtani & \cellcolor{litAlqahtani}fasttext (10-fold CV) & \cellcolor{litAlqahtani}fasttext & \cellcolor{litAlqahtani}0.93 & \cellcolor{litAlqahtani}0.81 & \cellcolor{litAlqahtani}0.67 & \cellcolor{litAlqahtani}0.80 & \cellcolor{litAlqahtani}0.77 & \cellcolor{litAlqahtaniSummary}0.80 & \cellcolor{litAlqahtaniSummary}0.08 \\
\cellcolor{litFranaEtAl}França et al. & \cellcolor{litFranaEtAl}GPT zero-shot & \cellcolor{litFranaEtAl}Falcon & \cellcolor{litFranaEtAl}- & \cellcolor{litFranaEtAl}0.55 & \cellcolor{litFranaEtAl}0.49 & \cellcolor{litFranaEtAl}0.45 & \cellcolor{litFranaEtAl}0.48 & \cellcolor{litFranaEtAlSummary}0.49 & \cellcolor{litFranaEtAlSummary}0.03 \\
\cellcolor{litFranaEtAl}França et al. & \cellcolor{litFranaEtAl}GPT zero-shot & \cellcolor{litFranaEtAl}Instruct & \cellcolor{litFranaEtAl}- & \cellcolor{litFranaEtAl}0.61 & \cellcolor{litFranaEtAl}0.48 & \cellcolor{litFranaEtAl}0.44 & \cellcolor{litFranaEtAl}0.45 & \cellcolor{litFranaEtAlSummary}0.50 & \cellcolor{litFranaEtAlSummary}0.07 \\
\cellcolor{litFranaEtAl}França et al. & \cellcolor{litFranaEtAl}GPT zero-shot & \cellcolor{litFranaEtAl}Openorca & \cellcolor{litFranaEtAl}- & \cellcolor{litFranaEtAl}0.54 & \cellcolor{litFranaEtAl}0.53 & \cellcolor{litFranaEtAl}0.54 & \cellcolor{litFranaEtAl}0.60 & \cellcolor{litFranaEtAlSummary}0.55 & \cellcolor{litFranaEtAlSummary}0.03 \\
\cellcolor{litFranaEtAl}França et al. & \cellcolor{litFranaEtAl}GPT zero-shot & \cellcolor{litFranaEtAl}Wizard & \cellcolor{litFranaEtAl}- & \cellcolor{litFranaEtAl}0.50 & \cellcolor{litFranaEtAl}0.40 & \cellcolor{litFranaEtAl}0.47 & \cellcolor{litFranaEtAl}0.48 & \cellcolor{litFranaEtAlSummary}0.46 & \cellcolor{litFranaEtAlSummary}0.04 \\
\cellcolor{litFranaEtAl}França et al. & \cellcolor{litFranaEtAl}ML random 80/20 & \cellcolor{litFranaEtAl}RF & \cellcolor{litFranaEtAl}- & \cellcolor{litFranaEtAl}0.04 & \cellcolor{litFranaEtAl}0.59 & \cellcolor{litFranaEtAl}0.68 & \cellcolor{litFranaEtAl}0.62 & \cellcolor{litFranaEtAlSummary}0.48 & \cellcolor{litFranaEtAlSummary}0.26 \\
\cellcolor{litFranaEtAl}França et al. & \cellcolor{litFranaEtAl}ML random 80/20 & \cellcolor{litFranaEtAl}LR & \cellcolor{litFranaEtAl}- & \cellcolor{litFranaEtAl}0.34 & \cellcolor{litFranaEtAl}0.53 & \cellcolor{litFranaEtAl}0.66 & \cellcolor{litFranaEtAl}0.53 & \cellcolor{litFranaEtAlSummary}0.52 & \cellcolor{litFranaEtAlSummary}0.11 \\
\cellcolor{litFranaEtAl}França et al. & \cellcolor{litFranaEtAl}ML random 80/20 & \cellcolor{litFranaEtAl}SVM & \cellcolor{litFranaEtAl}- & \cellcolor{litFranaEtAl}0.06 & \cellcolor{litFranaEtAl}0.19 & \cellcolor{litFranaEtAl}0.64 & \cellcolor{litFranaEtAl}0.35 & \cellcolor{litFranaEtAlSummary}0.31 & \cellcolor{litFranaEtAlSummary}0.22 \\
\cellcolor{litOurs}\textbf{Ours} & \cellcolor{litOurs}\textbf{embed + LogReg} & \cellcolor{litOurs}\textbf{MPNet + LogReg (Elastic)} & \cellcolor{litOurs}\textbf{0.80} & \cellcolor{litOurs}\textbf{0.72} & \cellcolor{litOurs}\textbf{0.58} & \cellcolor{litOurs}\textbf{0.50} & \cellcolor{litOurs}\textbf{0.64} & \cellcolor{litOursSummary}\textbf{0.65} & \cellcolor{litOursSummary}\textbf{0.10} \\
\cellcolor{litOurs}\textbf{Ours} & \cellcolor{litOurs}\textbf{embed + LogReg} & \cellcolor{litOurs}\textbf{MiniLM-L12 + LogReg (Elastic)} & \cellcolor{litOurs}\textbf{0.74} & \cellcolor{litOurs}\textbf{0.66} & \cellcolor{litOurs}\textbf{0.63} & \cellcolor{litOurs}\textbf{0.60} & \cellcolor{litOurs}\textbf{0.63} & \cellcolor{litOursSummary}\textbf{0.65} & \cellcolor{litOursSummary}\textbf{0.05} \\
\hline
\end{tabular}%
}
\endgroup

\end{table}

Cross-distribution stability is high: G-measure standard deviation across the five projects ($\sigma \approx 0.102$) is comparable to FARSEC ($\sigma = 0.101$), the strongest within-project baseline from Wu et al.~\cite{wu2022data-quality}.
MiniLM-L12 + LogReg (Elastic), the most stable of our variants, achieves $\sigma = 0.046$, the lowest cross-project standard deviation of all reported methods, while having a mean G-measure of $0.65$, which matches MPNet + LogReg (Elastic)'s mean G-measure of  $0.649$.
G-measure across fine-tuned small LLMs from Soltaniani et al., which train on each project's own data, exhibits substantially higher standard deviation (BERT $\sigma = 0.211$, DistilBERT $\sigma = 0.194$, Qwen2.5B $\sigma = 0.238$), suggesting that per-project fine-tuning amplifies sensitivity to distributional shifts rather than suppressing it.

\subsection*{C3: Preliminary Evaluation}

We evaluated the pipeline (requirement extraction $\to$ bi-encoder recall $\to$ cross-encoder rerank $\to$ LLM semantic filter) through semi-structured interviews with industry practitioners, using internal security policies as the primary source corpus and CIS Benchmarks (nginx, docker-ce) as a secondary corpus to probe generalization to concrete best-practice security guidelines.
This evaluation served to provide preliminary performance evidence and to surface design requirements.

Two semi-structured practitioner interview sessions are reported. Together they covered $24$ scored clauses and recommendations, drawn from both corpora, across six backlog items.
Items were selected from the set flagged by C2 at confidence scores between $0.55$ and $0.93$, spanning high-confidence and low-confidence regions of the classifier's output, hence representative of what the pipeline would surface in practice.

\subsubsection*{Retrieval Quality}\label{sec:results-c3-retrieval}

The analysis of the retrieval quality is reported over the $24$ rated clauses, hence summarizing $24$ ratings from two raters over six backlog items.
They thus characterize the study sample and their role is to contextualize the qualitative findings and to highlight failure modes.
As presented in \Cref{tab:c3-rating-distribution}, ratings cluster bimodally: $12/24$ at $\geq 4/5$, $7/24$ at $\leq 1/5$, and $5/24$ in the $2$--$3$ band, a distribution consistent with, but not sufficient to confirm, a pipeline whose shortlists are confident and whose applicability is determined by item-security requirement granularity alignment rather than by rater noise.
Internal-policy clauses dominated the high-rated band on items with well-structured descriptions, while the CIS Benchmark shortlists showed greater within-shortlist standard deviation, concentrated on a single infrastructure item.

\begin{table}[h]
    \centering
    \small
    \caption{Distribution of Requirements Relevance Ratings Across the 24 Security Requirements Scored in the Interview Sessions.}
    \label{tab:c3-rating-distribution}
    \begin{tabular}{lcc}
        \hline
        Rating band & Count & Share \\
        \hline
        $\geq 4/5$ (directly applicable) & $12$ & $50\%$ \\
        $2$--$3/5$ (partial)             & $5$  & $21\%$ \\
        $\leq 1/5$ (poor/irrelevant)     & $7$  & $29\%$ \\
        \hline
    \end{tabular}
\end{table}

\textbf{Failure modes.}
Two failure modes account for the low-rated security requirements and are architecturally informative.
\emph{Surface-keyword mismatch.} A JWT-generation item retrieved a password-handling clause because both topics involve authentication credentials at the embedding level, but the clause's normative content (``passwords must not be logged'') is inapplicable to JWT issuance.

\emph{Shortlist heterogeneity.} On one infrastructure item the CIS Benchmark retrieval returned a strongly applicable recommendation (rated $5/5$) alongside weakly applicable recommendations on adjacent topics (rated $1/5$) on the same shortlist, with the internal-policy retrieval on the same item rated more precisely.

\subsubsection*{Top-ranked examples}\label{sec:results-c3-examples}

Two items illustrate the two most informative patterns in the case.

\emph{Item~A} --- C2 confidence $0.86$. A logging configuration task. Three out of four retrieved clauses were rated $\geq 4/5$. Practitioner commentary described them as directly usable without paraphrasing. The last clause was rated $2/5$ and was thus not described as relevant. This is the pipeline's best-case output: concrete task, rule-level corpus, precise lexical and normative overlap.

\emph{Item~B} --- C2 confidence $0.93$. A scan and event notification configuration task. Only one of the retrieved clauses from the internal policy was rated $\geq 4/5$, the rest of them were rated $\leq 2/5$ and described as not quite relevant to the source topic.

Both items lie in the top tercile of C2 confidence, matching the deployment scenario in which the tool would provide information to an engineer.

\subsubsection*{Practitioner Design Feedback}\label{sec:results-c3-feedback}

Valuable insights were gained from the semi-structured interview, which are reported here. These insights serve to justify future design decisions to integrate the tool into further workflows.

\textbf{In-backlog-system integration.}
Both practitioners identified integration into the existing issue tracker systems as the primary precondition for practical adoption.
A standalone interface adds tool-fatigue. Retrieval output surfaced as issue comments or side-panel content can be consumed at the point where security planning actually occurs.

\textbf{Security Requirements Verbosity.}
Retrieved security requirements from both corpora are often too long for inline reading.
The recommended mitigation, using the same LLM stage to generate a short, concrete action-item summary per retained requirement with a link back to the authoritative security requirement, would address this directly without modifying the retrieval architecture.

\textbf{Corpus coverage.}
The one recurring gap was that the source corpus occasionally omits cross-cutting requirements.
Backwards-compatibility obligations flagged in one session are not expressed as security requirements in any of the evaluated documents, confirming that the tool cannot surface what is not in the corpus.

\section{Discussion}
We interpret the results contribution by contribution: the design choices and scope of the C1 dataset, the classifier-design rationale and cross-distribution behavior of C2, and the architectural rationale, abstraction-level findings, and workflow-fit signals of C3.

\subsection*{C1: Dataset Quality and Scope}\label{sec:discussion-c1}

Wu et al.~\cite{wu2022data-quality} document substantial label noise in community-assigned security fields across the widely used Jira-based benchmarks, and subsequent classifier work~\cite{alqahtani2024fasttext, franca2025gpts, soltaniani2026evaluating-llms} inherits that noise even after Wu's cleaning pass.
C1 trades scale for annotation rigor: $288$ items fully multi-rated by $9$ security experts produce $972$ independent assessments at Fleiss' $\kappa = 0.787$.
The design target is to complement Wu et al.'s dataset with a high-precision training signal whose labels were assigned by practitioners whose day-to-day role is to decide exactly the kind of question the annotation asks.
Wu et al.'s datasets remain the larger-scale evaluation. C1 is the clean training supervision.

The $13.5\%$ of items with at least one dissenting assessor, despite substantial inter-rater agreement overall, confirm that security relevance is not a sharp classification boundary.
A natural generalizability question for any expert-annotated resource is whether a different expert pool would converge on similar labels.
The annotators share a common organizational context, but that context is broad due to the size of the industrial partner, which operates across multiple highly regulated domains and is therefore subject to a wide cross-section of security standards and regulations.
This implies that the practitioners' day-to-day decisions span much of the compliance landscape that other security-regulated organizations also face, making their experience representative of many industries rather than narrowly organization-specific.
The working definition of \texttt{security-relevant} that they apply is in turn shaped by the same public standards (e.g., IEC 62443, CIS Benchmarks) that those organizations follow, so it is not idiosyncratic to a single industry.

The source corpus is intentionally narrow: two Atlassian enterprise projects whose item language resembles the industrial backlogs targeted by C3.
Datasets whose item descriptions are tersely phrased or centered on consumer-facing bug reports sit at a different item granularity and are better served by the Wu family.
The $80$ security-related items also bound what future work can extract from C1 alone: frozen-encoder classifiers trained as in C2 reach the ceiling of this signal comfortably, but multi-class Firesmith-category models~\cite{firesmith2004specifying} would require augmentation or a second annotation round.

\paragraph{Using C1 in practice.}
Beyond its role as training supervision for C2, we see three concrete uses of the released dataset for industry practitioners.
First, as expert-labeled seed data to bootstrap an organization-internal security-relevance classifier before internal labels exist, following the C2 procedure of a frozen encoder with a lightweight head.
Second, as an audit set. Organizations already operating a triage classifier or LLM-based filter can evaluate it against expert consensus labels rather than community-assigned ones.
Third, as realistic material for security training. The items are industry-style backlog texts with expert consensus security-relevance labels, which teams can use to align their own working definition of security relevance.

\subsection*{C2: Classifier Design Decisions}\label{sec:discussion-c2}

With $288$ annotated items, full fine-tuning of a contemporary transformer risks overfitting to the two source projects rather than learning a transferable notion of security-relevance.
Linear-head coefficients are also inspectable: security-filtering decisions are reviewed internally, and an auditable decision path matters for trust and for error analysis.
F2 as the selection criterion encodes the corresponding cost asymmetry fairly, since a missed security requirement is more expensive than a false alarm that C3 subsequently filters through its semantic stage.

\subsubsection*{Generalizability evaluation}\label{sec:discussion-c2-generalizability}

The generalizability results should be interpreted as evidence of transferability.
The deployed C2 model is evaluated fully zero-shot across domains: trained only on C1 and then applied unchanged to the Wu et al.\ benchmarks, whereas most published baselines are evaluated in-distribution, with training and test data drawn from the same project.
Matching or approaching those baselines is therefore a comparatively demanding target rather than a neutral head-to-head comparison.

The comparison against prompted LLMs should also be read cautiously. Although they are evaluated in a nominally zero-shot setting, their pretraining data may plausibly already contain parts of these widely used benchmark projects, potentially resulting in overly optimistic results~\cite{deng2024llm-contamination, baltes2026guidelinesempiricalstudiessoftware}. We therefore treat competitive performance on Wu et al.\ as a positive result showing that C2 has learned a transferable notion of \texttt{security-relevant}.

The reported superior performance of Gemini by Soltaniani et al.~\cite{soltaniani2026evaluating-llms}, while impressive, cannot be perceived as a stable baseline given recent insights on the non-reproducibility of research performed with commercial LLMs~\cite{angermeirReproducibility}.

The tight clustering of in-distribution F2 across eight sentence-encoder-based combinations (F2 $\in [0.625, 0.824]$) suggests that the task is well-posed and recoverable by any competitive text representation. The bottleneck here is label quality.
DistilRoBERTa + LogReg (L2) has the highest held-out F2 overall, but was not treated as a deployment candidate because on a two-project training resource its $\ell_2$ head is more exposed to project-specific overfitting than the sparser elastic-net alternatives.
Among the transfer-oriented deployment candidates, MPNet + LogReg (Elastic) has the highest held-out F2 and was therefore selected for integration into the C3 retrieval system.
The cross-distribution comparison uses mean G-measure as its primary transfer metric and across-project standard deviation $\sigma$ as a secondary stability metric.
Under that criterion, MPNet + elastic-net logistic regression remains a strong deployment choice, but the stability comparison also surfaces a real trade-off: MiniLM-L12 + elastic-net logistic regression achieves essentially identical mean G-measure ($0.650$ vs.\ $0.649$) with half the cross-project standard deviation ($\sigma = 0.046$ vs.\ $0.102$).
The fact that both variants are competitive with the strongest within-project baselines on stability, and that fine-tuned small LLMs are considerably less consistent despite per-project training supervision, supports the interpretation that the classifier's stability derives from the transferability of the learned representation.
In deployments where cross-domain consistency is the primary concern, MiniLM-L12 + LogReg (Elastic) would therefore be the preferred variant.

Probability calibration is deliberately deferred: C2 is used downstream as a hard filter into C3, where calibration does not affect retrieval-pipeline behavior.
This would matter only if C2 output fed a downstream ranking, which is not the case in the current deployment.

\subsection*{C3: Security Requirement Enrichment}\label{sec:discussion-c3}

We discuss C3 along its design rationale, the practical value to developer workflows observed in the preliminary evaluation, and four findings that emerged in individual walkthroughs.

\subsubsection*{Design rationale}

\emph{Two-stage C2--C3 system.}
Separating the high-recall relevance filter from the precision-oriented retrieval stage matches the asymmetric cost of each stage and lets each subsystem be tuned independently: C2 maximizes recall of security-relevant items, C3 maximizes precision of clause grounding.
The LLM workload is therefore proportional to the number of security-relevant items, not to the full backlog size.

\emph{RAG over pure LLM IR.}
The bi-encoder shortlist imposes a document-grounded candidate set that the LLM filter cannot hallucinate out of, while the LLM filter rejects topically adjacent but non-applicable candidates that dense similarity alone cannot distinguish.

\emph{Requirement extraction preprocessing.}
Introduced after we observed a systematic failure mode where multi-concern backlog items diluted the retrieval signal of any single concern.
Decomposing into atomic requirements before retrieval recovers concerns that the undecomposed retrieval pipeline misses.

\emph{Open-weights LLMs.}
Using on-premises open-weights models for both extraction and filtering satisfies two deployment constraints simultaneously: backlog contents cannot be sent to third-party inference services, and pinning a specific model version prevents silent behavior drift as commercial models are updated. More of the benefits of open LLMs can be found in the LLM guidelines by Baltes et al.~\cite{baltes2026guidelinesempiricalstudiessoftware}.

\subsubsection*{Practical value}

The tool surfaces implicit security requirements that would otherwise require manual expert review, delivering security requirements at the point of planning rather than as an after-the-fact compliance check, potentially addressing the lead-time burden identified by Moyón et al.~\cite{moyon2024challenges}.
In the preliminary evaluation, both practitioners responded positively to this framing. The tool was perceived as useful at the item level, provided it integrates into existing workflows rather than introducing a parallel interface.

\subsubsection*{Abstraction-level mismatch}\label{sec:discussion-c3-abstraction}

Across both interview sessions, the granularity gap between security requirement documents and backlog items appeared to be the strongest predictor of retrieval usefulness.
Rule-level documents, such as organizational security policies and implementation-level benchmarks such as CIS Benchmarks, operate below the capability abstraction of standards such as IEC 62443-3-3~\cite{iec2013iec62443-3-3}, and yielded actionable matches in both cases.
However, the within-shortlist variance of CIS Benchmark retrievals was higher, consistent with their narrower implementation scope making them precise for specific infrastructure items but less broadly applicable across item types.
Higher-abstraction capability standards, such as IEC 62443-3-3, produced retrievals that were topically correct but rarely directly actionable per item. The semantic gap between such control and a backlog item describing a specific feature is too large for retrieval to bridge reliably.
If confirmed by the upcoming complete evaluation, this finding has a direct implication for deployment.
The retrieval pipeline should be paired with documents whose security requirements granularity matches the typical specificity of the target backlog.

A secondary observation is that retrieval shortlist quality appears to track backlog-item granularity itself. Coarse epics retrieve many weakly relevant security requirements, while atomic tasks retrieve fewer but more strongly relevant ones.
This suggests a potential secondary use of C2\,+\,C3 output as a proxy signal for backlog-structuring quality, though we do not validate this claim in this paper.

\subsubsection*{Workflow fit}

Both practitioners independently identified in-backlog-system integration as the primary precondition for practical adoption.
Standalone interfaces were perceived to add tool-fatigue. Retrieval output surfaced as issue comments or side-panel content can be consumed at the point where planning actually happens.
This is considered the most consistent qualitative finding of the interview study.

Security requirement verbosity was the dominant usability concern. Retrieved clauses from both internal policies and public benchmarks are often too long for inline reading.
A natural next engineering step is to use the same LLM stage to generate a short, concrete action-item summary per retained security requirement, with a link back to the authoritative text for verification, an addition that has been implemented but must be evaluated.
Perceived usefulness also appeared to correlate with source-item quality. Items with structured descriptions produced higher-rated retrievals than items with terse, unstructured titles.

\subsubsection*{Missing items and concerns}

The interviews produced few systematic gap reports.
The one recurring finding was that the source corpus occasionally omits cross-cutting requirements. Backwards-compatibility obligations flagged in one session are not expressed as security clauses in any of the evaluated standards, and the tool cannot surface what is not in the corpus.
The extractor continues to have difficulties with multi-concern items, which is why further multi-concern decomposition is an open direction.
This limitation may also be mitigated by establishing guidelines for backlog item descriptions, since more detailed items may provide better source material for retrieving relevant security requirements.

\subsubsection*{Generalizability}
The C2 classifier is trained entirely on open-source TAWOS data and carries no organization-specific knowledge, so adopters applying C3 to their own internal security policies need no re-annotation or retraining.
The strongest generalizability claim for C3, that the retrieval pipeline performs consistently across standards and organizations, is part of future work.
The CIS Benchmark evaluation partially addresses this by demonstrating consistent pipeline behavior on a public security requirements document against the same items used for the internal-policy evaluation, but a second independent industrial corpus would be needed to convert the architectural document-agnosticism claim into a measured result.

\section{Threats to Validity}
We discuss the threats to validity of our research following the categorization of Runeson and Höst~\cite{Runeson2009}.
\subsection{Construct Validity}

The construct \emph{security-relevance} is operationalized through expert judgment rather than as an approximation of an objective ground truth. 
Different expert pools might apply a different decision boundary, and the $13.5\%$ of C1 items with at least one disagreeing assessor confirm that the construct admits legitimate boundary cases, as discussed in ~\Cref{sec:discussion-c1}.
We mitigate this by multi-rating every released item, and by reporting Fleiss' $\kappa$ and raw observed agreement rather than a single reliability number.
For C3, \emph{retrieval usefulness} is characterized in two layers: an artifact-level analysis of shortlists (applicability relative to the retrieving item's requirement) and a practitioner $0$--$5$ rating used as calibration of that analysis.

\subsection{Internal Validity}

C2 thresholds and hyperparameters are selected on a single C1 validation split.
The tight spread of F2 across the sentence-encoder variants bounds the risk of threshold-to-split overfitting but does not eliminate it, and probability calibration is not performed beyond threshold choice because C2 is used downstream as a hard filter into C3.
The held-out C1 test set contains only $58$ items, so the reported in-distribution point estimates probably carry wide confidence intervals and, taken alone, cannot rule out overfitting to the two source projects.
We mitigate this by not relying on the in-distribution split for the headline claim. Deployment selection is restricted to the transfer-oriented candidate set, and the zero-shot evaluation on the five Wu et al.\ benchmarks~\cite{wu2022data-quality}, spanning several thousand items from unrelated projects, provides the primary evidence that the learned notion of security relevance generalizes.
The C3 evaluation is preliminary and performed over two industrial product backlogs and two practitioner interviews.
Observations are descriptive and the apparent convergences across the two sessions are consistent with, but do not establish generalizable empirical evidence.
The complete evaluation, involving more practitioners per backlog, multiple backlogs, ideally across organizations, and a study design capable of measuring workflow impact rather than perceived relevance, is deferred to future work.
Item-level familiarity is recorded as descriptive context.
The interviews target the engineer-consulting scenario in which the consumer of a security requirement is not necessarily the author of the corresponding backlog item.
Prior familiarity with a specific backlog item does not, however, disqualify a practitioner's judgment of whether a retrieved security requirement is applicable.

\subsection{External Validity}

C1 is drawn from two Atlassian open-source projects whose item language resembles industrial enterprise backlogs. It is a precision resource by design and does not aim to cover the broader variance of public issue trackers (Wu et al.'s benchmarks fill that role).
The nine C1 annotators share an organizational context, which could bias the operational definition of \texttt{security-relevant}.
We argue in Section~\ref{sec:discussion-c1} that this context is operationally shaped by the same standards other security-regulated organizations follow.

\subsection{Reliability}

Both the C3 requirement extractor and the semantic filter are open-weights LLMs, and both stages introduce run-to-run variability in the resulting shortlist.
We mitigate this by fixing the model identifier, sampling temperature, and prompts, and by reporting averaged behavior.
We do not claim bitwise reproducibility of the C3 output.
Although we do not quantify run-to-run variance formally, repeated executions during development informally suggested that the C3 pipeline was practically consistent: for the same backlog item and source corpus, it reliably returned the same clauses and only small differences in relevance scores.
C1 and the full C2 pipeline are fully reproducible from the released artifacts and are deterministic up to library-level floating-point variation.
The C3 cross-distribution probe on CIS Benchmarks is reproducible end-to-end against public documents. The internal-policy evaluation is reproducible only on the internal corpus.

\section{Conclusion}

We set out to reduce the security workload in continuous security compliance by automatically including security requirements on the backlog items engineers and security experts already use.
The three contributions of this paper address complementary parts of that goal.

C1 provides an independently expert-annotated backlog-security dataset: $288$ items annotated by a pool of nine security practitioners, resulting in $972$ independent binary assessments, Fleiss' $\kappa = 0.787$, and additional Firesmith security-requirement metadata available for many items.
C1 is positioned as a high-precision complement to the larger Wu et al.\ benchmark family~\cite{wu2022data-quality}, and is released publicly to allow downstream work on security-relevance detection to train and audit against expert supervision.

C2 is a recall-oriented binary classifier for security-relevance.
A frozen MPNet encoder paired with an elastic-net logistic-regression head reaches F2 $=0.774$ on the C1 held-out split and, applied zero-shot to the five Wu cleaned benchmarks, achieves cross-distribution G-measure $\approx 0.65$, competitive with the published classical-ML and open-source-GPT baselines despite a prevalence shift from $27.8\%$ on C1 to $1.9$--$17.9\%$ on the Wu projects.

C3 is a four-stage retrieval-augmented generation pipeline that maps a flagged backlog item to the subset of security requirements from relevant security documents.
The C3 retrieval pipeline was evaluated preliminarily on two industrial product backlogs with industrial internal policies as the primary corpus and CIS Benchmarks as a public-standard probe.
Semi-structured practitioner interviews produced a positive preliminary signal on items with well-structured descriptions, identified integration into backlog systems and security requirement summarization as the most important additions, and suggested a consistent lesson about the granularity gap between security requirement documents and backlog items.
Rule-level documents yield more precise, actionable matches than higher-abstraction capability standards.

Taken together, the results provide preliminary evidence that security-relevance can be decided directly from ordinary backlog text to support engineering work through concrete security requirements attached to those items in a way that potentially makes security both more visible and better integrated in the software engineering lifecycle, without requiring engineers to leave their existing workflow.

~\label{sec:future-work}
Open directions follow directly from the preliminary evaluation.
First, a complete evaluation of C3 is required, including more practitioner interviews, multiple backlogs, ideally across organizations, and a controlled study design capable of measuring workflow impact and adoption.
This would be achieved through a longitudinal study.
Second, applying C3 to a second industrial corpus, and ideally to a second organization's policies, would convert the current document-agnostic architectural claim into a measured generalizability result.
Third, in-tracker integration and per-security requirements action-item summarization are concrete engineering steps that the practitioners identified as the dominant preconditions for practical adoption.

\section*{Data Availability Statement}
We provide the artifacts of this research in two parts: the final C1 dataset covering all 288 items with aggregated labels and associated metadata~\cite{garcia2026dataset}, and the code and model artifacts for C2 and the sanitized C3 retrieval pipeline~\cite{garcia2026pipeline}.
Due to confidentiality agreements, we do not release the segmenter for the internal policy documents or the internal policies.

% \bibliographystyle{ACM-Reference-Format}
% \bibliography{main.bib}

\printbibliography

@article{wu2022data-quality,
  author={Wu, Xiaoxue and Zheng, Wei and Xia, Xin and Lo, David},
  journal={IEEE Transactions on Software Engineering}, 
  title={Data Quality Matters: A Case Study on Data Label Correctness for Security Bug Report Prediction}, 
  year={2022},
  volume={48},
  number={7},
  pages={2541-2556},
  doi={10.1109/TSE.2021.3063727}
}

@misc{soltaniani2026evaluating-llms,
      title={Evaluating Large Language Models for Security Bug Report Prediction}, 
      author={Farnaz Soltaniani and Shoaib Razzaq and Mohammad Ghafari},
      year={2026},
      eprint={2601.22921},
      archivePrefix={arXiv},
      primaryClass={cs.CR},
      url={https://arxiv.org/abs/2601.22921}, 
}

@inproceedings{moyon2024challenges,
    author = {Moy\'{o}n, Fabiola and Angermeir, Florian and Mendez, Daniel},
    title = {Industrial Challenges in Secure Continuous Development},
    year = {2024},
    isbn = {9798400705014},
    publisher = {Association for Computing Machinery},
    address = {New York, NY, USA},
    url = {https://doi.org/10.1145/3639477.3639736},
    doi = {10.1145/3639477.3639736},
    booktitle = {Proceedings of the 46th International Conference on Software Engineering: Software Engineering in Practice},
    pages = {309–311},
    numpages = {3},
    location = {Lisbon, Portugal},
    series = {ICSE-SEIP '24}
}

@inproceedings{moyon2020security-compliance-agile,
  author={Moyón, Fabiola and Almeida, Pamela and Riofrío, Daniel and Mendez, Daniel and Kalinowski, Marcos},
  booktitle={2020 46th Euromicro Conference on Software Engineering and Advanced Applications (SEAA)}, 
  title={Security Compliance in Agile Software Development: A Systematic Mapping Study}, 
  year={2020},
  pages={413-420},
  doi={10.1109/SEAA51224.2020.00073}
}

@article{alqahtani2024fasttext,
  author = {Alqahtani, Sultan S.},
  title = {Security Bug Reports Classification Using {fastText}},
  journal = {International Journal of Information Security},
  volume = {23},
  number = {2},
  pages = {1347--1358},
  year = {2024},
  doi = {10.1007/s10207-023-00793-w}
}

@article{franca2025gpts,
  author = {Fran\c{c}a, Higor and Goseva-Popstojanova, Katerina and Teixeira, C\'{a}ssio and Laranjeiro, Nuno},
  title = {{GPT}s Are Not the Silver Bullet: Performance and Challenges of Using {GPT}s for Security Bug Report Identification},
  journal = {Information and Software Technology},
  volume = {185},
  pages = {107778},
  year = {2025},
  doi = {10.1016/j.infsof.2025.107778}
}

@article{sangaroonsilp2023privacy-issue-reports,
  author = {Sangaroonsilp, Pattaraporn and Choetkiertikul, Morakot and Dam, Hoa Khanh and Ragkhitwetsagul, Chaiyong and Ghose, Aditya},
  title = {An Empirical Study of Automated Privacy Requirements Classification in Issue Reports},
  journal = {Automated Software Engineering},
  volume = {30},
  number = {20},
  year = {2023},
  doi = {10.1007/s10515-023-00387-9}
}

@inproceedings{ayala-rivera2018grace-period,
  author = {Ayala-Rivera, Vanessa and Pasquale, Liliana},
  title = {{The Grace Period Has Ended}: An Approach to Operationalize {GDPR} Requirements},
  booktitle = {2018 IEEE 26th International Requirements Engineering Conference (RE)},
  year = {2018},
  pages = {136--146},
  doi = {10.1109/RE.2018.00023}
}

@article{ayala-rivera2024soco,
  author = {Ayala-Rivera, Vanessa and Portillo-Dominguez, Andres Omar and Pasquale, Liliana},
  title = {{GDPR} Compliance via Software Evolution: Weaving Security Controls in Software Design},
  journal = {Journal of Systems and Software},
  volume = {216},
  pages = {112144},
  year = {2024},
  doi = {10.1016/j.jss.2024.112144}
}

@article{ruiz2023traceability,
  author = {Ruiz, Marcela and Hu, Jin Yang and Dalpiaz, Fabiano},
  title = {Why Don't We Trace? A Study on the Barriers to Software Traceability in Practice},
  journal = {Requirements Engineering},
  volume = {28},
  pages = {619--637},
  year = {2023},
  doi = {10.1007/s00766-023-00408-9}
}

@inproceedings{deng2024llm-contamination,
    title = "Investigating Data Contamination in Modern Benchmarks for Large Language Models",
    author = "Deng, Chunyuan and
              Zhao, Yilun and
              Tang, Xiangru and
              Gerstein, Mark and
              Cohan, Arman",
    booktitle = "Proceedings of the 2024 Conference of the North American Chapter of the Association for Computational Linguistics: Human Language Technologies (Volume 1: Long Papers)",
    month = jun,
    year = "2024",
    address = "Mexico City, Mexico",
    publisher = "Association for Computational Linguistics",
    pages = "8706--8719",
    doi = "10.18653/v1/2024.naacl-long.482",
    url = "https://aclanthology.org/2024.naacl-long.482/"
}

@ARTICLE{raharjana2021user-stories-nlp,
  author={Raharjana, Indra Kharisma and Siahaan, Daniel and Fatichah, Chastine},
  journal={IEEE Access}, 
  title={User Stories and Natural Language Processing: A Systematic Literature Review}, 
  year={2021},
  volume={9},
  number={},
  pages={53811-53826},
  doi={10.1109/ACCESS.2021.3070606}
}

@inproceedings{ferreira2022writing-agile-requirements,
    author={António M. S. Ferreira and Alberto {Rodrigues da Silva} and Ana C. R. Paiva},
    title={Towards the Art of Writing Agile Requirements with User Stories, Acceptance Criteria, and Related Constructs},
    booktitle={Proceedings of the 17th International Conference on Evaluation of Novel Approaches to Software Engineering - ENASE},
    year={2022},
    pages={477-484},
    publisher={SciTePress},
    organization={INSTICC},
    doi={10.5220/0011082000003176},
    isbn={978-989-758-568-5},
    issn={2184-4895},
}

@manual{iso2018sse,
  title = {{ISO/IEC/IEEE} 26515:2018 Systems and software engineering -- Developing information for users in an agile environment},
  organization = {ISO/IEC/IEEE},
  year = {2018},
  number = {26515:2018},
}

@inproceedings{reimers2019sbert,
  author    = {Reimers, Nils and Gurevych, Iryna},
  title     = {Sentence-{BERT}: Sentence Embeddings using Siamese {BERT}-Networks},
  booktitle = {Proceedings of the 2019 Conference on Empirical Methods in Natural Language Processing},
  series    = {EMNLP '19},
  year      = {2019},
  pages     = {3982--3992},
  doi       = {10.18653/v1/D19-1410},
  publisher = {Association for Computational Linguistics}
}

@incollection{lewis2020rag,
  author    = {Lewis, Patrick and Perez, Ethan and Piktus, Aleksandra and Petroni, Fabio and Karpukhin, Vladimir and Goyal, Naman and K\"{u}ttler, Heinrich and Lewis, Mike and Yih, Wen-tau and Rockt\"{a}schel, Tim and Riedel, Sebastian and Kiela, Douwe},
  title     = {Retrieval-Augmented Generation for Knowledge-Intensive {NLP} Tasks},
  booktitle = {Advances in Neural Information Processing Systems},
  volume    = {33},
  pages     = {9459--9474},
  year      = {2020},
  publisher = {Curran Associates}
}

@misc{nogueira2019bert,
  author        = {Nogueira, Rodrigo and Cho, Kyunghyun},
  title         = {Passage Re-ranking with {BERT}},
  year          = {2019},
  eprint        = {1901.04085},
  archivePrefix = {arXiv},
  primaryClass  = {cs.IR}
}

@inproceedings{tawosi2022tawos,
author = {Tawosi, Vali and Al-Subaihin, Afnan and Moussa, Rebecca and Sarro, Federica},
title = {A versatile dataset of agile open source software projects},
year = {2022},
isbn = {9781450393034},
publisher = {Association for Computing Machinery},
address = {New York, NY, USA},
doi = {10.1145/3524842.3528029},
booktitle = {Proceedings of the 19th International Conference on Mining Software Repositories},
pages = {707–711},
numpages = {5},
series = {MSR '22}
}

@article{firesmith2004specifying,
  author  = {Firesmith, Donald G.},
  title   = {Specifying Reusable Security Requirements},
  journal = {Journal of Object Technology},
  volume  = {3},
  number  = {1},
  pages   = {61--75},
  year    = {2004},
  doi     = {10.5381/jot.2004.3.1.c6}
}

@article{zou2005elastic-net,
  author  = {Zou, Hui and Hastie, Trevor},
  title   = {Regularization and Variable Selection via the Elastic Net},
  journal = {Journal of the Royal Statistical Society: Series B (Statistical Methodology)},
  volume  = {67},
  number  = {2},
  pages   = {301--320},
  year    = {2005},
  doi     = {10.1111/j.1467-9868.2005.00503.x}
}

@misc{baltes2026guidelinesempiricalstudiessoftware,
      title={{Guidelines for Empirical Studies in Software Engineering involving Large Language Models}}, 
      author={Sebastian Baltes and Florian Angermeir and Chetan Arora and Marvin Muñoz Barón and Chunyang Chen and Lukas Böhme and Fabio Calefato and Neil Ernst and Davide Falessi and Brian Fitzgerald and Davide Fucci and Junda He and Christoph Treude and Marcos Kalinowski and Stefano Lambiase and Daniel Russo and Mircea Lungu and Cristina Martinez Montes and Lutz Prechelt and Paul Ralph and Rijnard van Tonder and Stefan Wagner},
      year={2025},
      eprint={2508.15503},
      archivePrefix={arXiv},
      primaryClass={cs.SE},
      url={https://arxiv.org/abs/2508.15503}, 
}

@inproceedings{angermeirReproducibility,
  author    = {Angermeir, Florian and Amougou, Maximilian and Kreitz, Mark and Bauer, Andreas and Linhuber, Matthias and Fucci, Davide and Moyón C., Fabiola and Mendez, Daniel and Gorschek, Tony},
  title     = {Reflections on the Reproducibility of Commercial {LLM} Performance in Empirical Software Engineering Studies},
  booktitle = {IEEE/ACM
48th International Conference on Software Engineering},
  series    = {ICSE '26},
  year      = {2026},
  doi       = {10.48550/arXiv.2510.25506},
  publisher = {Association for Computing Machinery},
  address   = {New York, NY, USA}
}

@inproceedings{turpe2017scrum,
  title={Managing Security Work in Scrum: Tensions and Challenges},
  author={Sven T{\"u}rpe and Andreas Poller},
  booktitle={SecSE@ESORICS},
  year={2017},
  url={https://api.semanticscholar.org/CorpusID:4933742}
}

@article{fitzgerald2017continuous,
    title = {Continuous software engineering: A roadmap and agenda},
    journal = {Journal of Systems and Software},
    volume = {123},
    pages = {176-189},
    year = {2017},
    issn = {0164-1212},
    doi = {https://doi.org/10.1016/j.jss.2015.06.063},
    url = {https://www.sciencedirect.com/science/article/pii/S0164121215001430},
    author = {Brian Fitzgerald and Klaas-Jan Stol}
}

@techreport{eu2024cra,
  title        = {EU Cyber Resilience Act},
  author       = {{European Parliamentary Research Service}},
  institution  = {European Parliament},
  year         = {2022},
  url          = {https://www.europarl.europa.eu/thinktank/en/document/EPRS_BRI%282022%29739259},
  note         = {Briefing, European Parliamentary Research Service},
}

@techreport{iec2018secure,
  title        = {Security for Industrial Automation and Control Systems -- Part 4-1: Secure Product Development Lifecycle Requirements},
  author       = {{International Electrotechnical Commission}},
  institution = {International Electrotechnical Commission},
  year         = {2018},
  number       = {IEC 62443-4-1:2018},
  publisher    = {International Electrotechnical Commission},
  url          = {https://webstore.iec.ch/publication/33615}
}

@inproceedings{angermeir2024automatedcsc,
author = {Angermeir, Florian and Fischbach, Jannik and Moy\'{o}n, Fabiola and Mendez, Daniel},
title = {Towards Automated Continuous Security Compliance},
year = {2024},
isbn = {9798400710476},
publisher = {Association for Computing Machinery},
address = {New York, NY, USA},
url = {https://doi.org/10.1145/3674805.3690748},
doi = {10.1145/3674805.3690748},
booktitle = {Proceedings of the 18th ACM/IEEE International Symposium on Empirical Software Engineering and Measurement},
pages = {440–446},
numpages = {7},
location = {Barcelona, Spain},
series = {ESEM '24}
}

@inproceedings{moyon2018towardcsc,
author = {Moyón, Fabiola and Beckers, Kristian and Klepper, Sebastian and Lachberger, Philipp and Bruegge, Bernd},
title = {Towards continuous security compliance in agile software development at scale},
year = {2018},
isbn = {9781450357456},
publisher = {Association for Computing Machinery},
address = {New York, NY, USA},
url = {https://doi.org/10.1145/3194760.3194767},
doi = {10.1145/3194760.3194767},
booktitle = {Proceedings of the 4th International Workshop on Rapid Continuous Software Engineering},
pages = {31–34},
numpages = {4},
location = {Gothenburg, Sweden},
series = {RCoSE '18}
}

@article{Runeson2009,
	title        = {Guidelines for conducting and reporting case study research in software engineering},
	author       = {Per Runeson and Martin H{\"o}st},
	year         = 2009,
	journal      = {Empirical Software Engineering},
	publisher    = {Springer},
	volume       = 14,
	number       = 2,
	pages        = {131--164},
	doi          = {10.1007/s10664-008-9102-8}
}

@article{artstein2008irr,
    title = "Survey Article: Inter-Coder Agreement for Computational Linguistics",
    author = "Artstein, Ron  and
      Poesio, Massimo",
    journal = "Computational Linguistics",
    volume = "34",
    number = "4",
    year = "2008",
    url = "https://aclanthology.org/J08-4004/",
    doi = "10.1162/coli.07-034-R2",
    pages = "555--596"
}

@manual{iec2013iec62443-3-3,
  title        = {Industrial communication networks -- Network and system security -- Part 3-3: System security requirements and security levels},
  organization = {International Electrotechnical Commission},
  number       = {IEC 62443-3-3:2013},
  year         = {2013},
  address      = {Geneva, Switzerland}
}

@article{tantithamthavorn2020rebalancing,
  author={Tantithamthavorn, Chakkrit and Hassan, Ahmed E. and Matsumoto, Kenichi},
  journal={IEEE Transactions on Software Engineering}, 
  title={The Impact of Class Rebalancing Techniques on the Performance and Interpretation of Defect Prediction Models}, 
  year={2020},
  volume={46},
  number={11},
  pages={1200-1219},
  doi={10.1109/TSE.2018.2876537}
  }

@misc{cis2024benchmarks,
  author       = {{Center for Internet Security}},
  title        = {{CIS Benchmarks}},
  year         = {2024},
  howpublished = {\url{https://www.cisecurity.org/cis-benchmarks/}},
}

@misc{garcia2026dataset,
  author       = {García Núñez, Ignacio and Angermeir, Florian and Moyón Constante, Fabiola},
  title        = {{C1: An Expert-Annotated Dataset of Security-Relevant Product Backlog Items}},
  year         = {2026},
  month        = jul,
  howpublished = {Zenodo},
  doi          = {10.5281/zenodo.21450745},
  url          = {https://doi.org/10.5281/zenodo.21450745},
  note         = {Dataset, version 1.0.0}
}

@misc{garcia2026pipeline,
  author       = {García Núñez, Ignacio and Angermeir, Florian and Moyón Constante, Fabiola},
  title        = {{C2/C3: Security-Relevance Classifier and Security-Requirements Retrieval Pipeline for Product Backlogs}},
  year         = {2026},
  month        = jul,
  howpublished = {Zenodo},
  doi          = {10.5281/zenodo.21450906},
  url          = {https://doi.org/10.5281/zenodo.21450906},
  note         = {Software, version 1.0.0}
}

\end{document}